# Precision-informed Fourier compression of refractive index tomograms with voxelwise error control

**Dohyeon Lee[a,b], Juheon Lee[a,b], Sehyeon Lee[a,b], Juyeon Park[a,b], YongKeun Park[a,b,c*]**

[a]Department of Physics, Korea Advanced Institute of Science and Technology (KAIST), Daejeon 34141, Republic of Korea

[b]KAIST Institute for Imaging, KAIST, Daejeon 34141, Republic of Korea

[c]Tomocube Inc., Daejeon 34109, Republic of Korea

**Abstract**

Significance: Holotomography (HT) generates large refractive index (RI) volumes and time series, creating a need for compact storage with controlled numerical fidelity.

Aim: We sought to reduce storage by directly encoding reconstructed three-dimensional RI volumes in Fourier space with voxelwise validation against an assigned error tolerance.

Approach: Quantized Fourier coefficients represent volumetric structure, while spatial residual quantization uses a 1% safety margin without storing separate corrections for individual voxels. We evaluated cellular and organoid data, concurrent processing, and extensions to dielectric-tensor and absorption tomography.

Results: At a tolerance of $10^{-4}$ RI units, complete packets achieved compression ratios of 7.89–9.50 relative to float32 arrays and were 28–30% smaller than BitRound–Pcodec packets satisfying the same error limit. Concurrent processing reduced encoding and decoding wall times by factors of 5.14 and 4.81 without changing packets. The fixed margin passed all primary-setting tests but failed at one smaller tolerance in both precision-sweep inputs.

Conclusions: Precision-informed volumetric Fourier coding reduces storage of holotomograms with verified voxelwise fidelity at the primary tolerance, at additional computational cost.

*Correspondence: YongKeun Park, yk.park@kaist.ac.kr

Statement of Discovery: In holotomography (HT), direct encoding of reconstructed three-dimensional refractive index (RI) volumes in Fourier space reduces storage with a residual safety margin and voxelwise error validation. The tested primary setting satisfies a $10^{-4}$ RI-unit tolerance without storing additional voxel replacement values. The framework also accommodates dielectric-tensor and absorption tomograms through quantity-specific error criteria.

## Introduction

Holotomography (HT) reconstructs three-dimensional refractive index (RI) distributions for label-free analysis of cells and tissues[1,2]. High-resolution RI imaging generates large datasets that must be stored and recovered for quantitative analysis. Label-free contrast enables repeated observation over extended periods, including large area tissue imaging, long-term organoid imaging, and four-dimensional measurements of intracellular dynamics[3–8]. Storing a volume at each time point expands the data burden as imaging progresses from individual cells to organoids and tissues. Faster scanning and illumination multiplexing[9,10] further increase the amount of RI data to store, transfer and recover.

RI compression must reduce stored bytes while limiting changes to the quantitative voxel values. Chunked formats such as OME-Zarr support scalable access to multidimensional bioimaging arrays[11,12], while the numerical representation and codec determine each chunk's size and recovery cost. Lee et al. established HT compression baselines and showed that compression ratio and processing bandwidth must be considered alongside transfer bandwidth[13]. Bit Grooming reduces stored precision to improve lossless compressibility[14]; precision reduction followed by numerical compression, as in BitRound–Pcodec, provides a direct storage-saving pipeline[15,16]. Digit Rounding similarly reduces mantissa precision before lossless coding but preserves significant decimal digits rather than a fixed absolute RI-error tolerance[17]. Encoding correlations across all three spatial dimensions offers a complementary way to reduce storage while controlling each voxel's numerical error.

A compression tolerance expressed in RI units limits the numerical change introduced into a reconstruction, separately from the accuracy of the original measurement. Metrological assessment of HT distinguishes reconstructed RI fidelity from spatial resolution and other imaging characteristics[18]. Where independently characterized, system precision can inform the allowed change, while compression error is evaluated relative to the supplied reconstruction. Measurement uncertainty has similarly motivated precision reduction in cryo-electron microscopy and tomography[19].

Scientific compressors already control numerical error through prediction, transforms and multilevel coding[20]. SZ combines prediction with error-controlled quantization, while zfp transforms floating-point arrays[21–23]; tensor decomposition[24] and MGARD[25] provide other multidimensional representations. SPERR combines wavelet coding with pointwise corrections[26], and the FFCz preprint considers joint spatial- and frequency-domain constraints[27]. Storage comparisons must count coefficients, residuals, voxel replacement data and metadata together at a verified voxelwise tolerance. Analysis of zfp further shows how errors at individual numerical stages propagate to its mode-specific bounds[28].

We directly encode reconstructed RI tomograms in three-dimensional Fourier space and validate decoded voxels against an absolute RI-error tolerance. Quantized Fourier coefficients represent volumetric structure, and spatial residuals correct local errors using a quantization interval of 1.98ε without storing an additional list of voxel indices and replacement values for errors exceeding the tolerance. We evaluate complete-packet storage and computational cost against local BitRound–Pcodec and zfp implementations on public HeLa and organoid data under matched fidelity criteria. Experiments varying coefficient spacing examine how storage is divided between coefficients and residuals and the numerical limits of the fixed margin, while concurrent processing tests time-lapse throughput with unchanged independent packets. Extensions to incoherent dielectric tensor tomography (iDTT)[29] and quantitative absorption tomography (QAT) assess component-specific coding and error criteria[30].

## Methods

### *Reconstructed Tomography Data and Sampling*

The compression inputs were reconstructed RI tomograms. HT reconstruction rests on inverse-scattering theory and diffraction-based backpropagation[31,32], using either white-light illumination with axially scanned image stacks[33,34,35] or coherent illumination with a scanned incidence angle[36,37].

Public HeLa and organoid RI arrays were obtained from the BioImage Archive dataset S-BIAD1739 associated with the HT compression benchmark of Lee et al.[13,38] (Supplementary Methods S1). The saved HeLa and organoid arrays used as native references in this study were float64. We retained those arrays as the numerical references and used their float32 casts as working inputs. Each working subvolume contained 2,097,152 voxels, corresponding to 8 MiB. Supplementary Methods S10 specifies the reproduction materials and remaining validation limits.

For the time-lapse experiment, we extracted 16 consecutive organoid subvolumes at archive time indices 75–90, using the same spatial region at every time point. Each subvolume contained $32 \times 256 \times 256$ voxels. The spatial location was selected from the native morphology before timing and held fixed across all time points.

The extension experiments used one region of interest (ROI) from iDTT and one melanoma ROI from QAT[29,30]. The iDTT and QAT data used in this study will be deposited in the BioImage Archive.

*Compression Tolerance and Numerical Reference*

Let $n_0(v)$ denote the native reconstructed RI at voxel $v$, $x(v) = f32(n_0(v))$ the float32 working value, and $\hat{n}(v)$ the decoded RI. Here, f32 and f64 denote casting to float32 and float64, respectively. We evaluated the maximum absolute RI error against an assigned tolerance $\varepsilon$, accepting a setting only if

$$\delta n_{max} = \max_{v} |\hat{n}(v) - n_0(v)| \leq \varepsilon .$$

The primary $\varepsilon = 10^{-4}$ RI-unit setting was the compression tolerance adopted in this study, selected with reference to our experimental measurements using an HT-X1 system (Tomocube Inc., Republic of Korea). This tolerance is defined relative to the supplied reconstruction; it is not a claim of absolute RI accuracy or a universal instrument specification. Different physical quantities in the extension experiments used the separate criteria specified below. We chose a maximum-error criterion to limit every voxel, including rare local changes that an average error can conceal. Root-mean-square error (RMSE) is also reported as a complementary metric. An orthonormal transform preserves squared error in exact arithmetic, but small transform-domain RMSE alone does not ensure a voxelwise maximum-error bound.

*Fourier Coefficient Representation and Lossless Coding*

For a working volume $x$ containing $N$ voxels, the mean $\mu$ and Fourier coefficients $F$ were computed in float64 as

$$\mu = \frac{1}{N} \sum_{v} x(v), F = F_r\left(f64(x) - \mu\right),$$

where $F_r$ denotes an orthonormal three-dimensional real fast Fourier transform (FFT). The stored half-spectrum had shape $(N_z, N_y, \lfloor N_x/2 \rfloor + 1)$, where $N_z$, $N_y$ and $N_x$ are the voxel counts along the three array axes. At frequency-grid index $k$, integer arrays $q_R$ and $q_I$ were obtained by independently quantizing the real and imaginary parts of $F$ with coefficient interval $\Delta_F$. The function *rint* rounds to the nearest integer with ties to even:

$$q_R(k) = rint\left(\frac{\mathrm{Re}\,F(k)}{\Delta_F}\right), q_I(k) = rint\left(\frac{\mathrm{Im}\,F(k)}{\Delta_F}\right).$$

Quantized integer coefficients are converted to floating-point values for dequantization and inverse FFT computation. The inverse-transform prediction is computed in float64 and cast to float32 only after reconstruction and mean addition. The integer storage width is independent of the floating-point precision used for reconstruction. The primary HT setting was $\Delta_F = 5\varepsilon$, an empirical rule selected during method development and fixed before the HeLa/organoid comparison. The factor 5 is an empirically chosen value held constant across the comparison, not a universal optimum or a chunk-size scaling law. Figure 4c,d gives the tested size trade-off: the minima differ between HeLa and organoid. The orthonormal FFT convention avoids the linear-in-voxel-count amplitude factor of an unnormalized discrete Fourier transform, but packet efficiency can still depend on chunk shape and boundary structure.

Quantized coefficients were partitioned into four radial bands to form separate numerical coding streams. For per-axis FFT frequencies $f_z, f_y, f_x$ in cycles per voxel, the normalized radius $\rho(k)$ was

$$\rho(k) = 2\sqrt{f_z^2 + f_y^2 + f_x^2}.$$

Boundaries at 0.25, 0.5 and 1.0 defined four bands, with equality assigned to the higher band. The metric uses per-axis Nyquist-normalized digital frequencies; it does not represent physical optical-transfer support. All half-spectrum positions were represented, and no Ewald-volume or resolution-based hard mask was applied. All four bands use the same coefficient interval $\Delta_F$.

Quantized real and imaginary arrays were cast to the smallest supported signed integer width, 16, 32 or 64 bits, covering their full ranges; all frequency bands used that same integer width. The integer storage width for the spatial residual was chosen separately. Each stream was losslessly encoded with Pcodec[16] at level 6. The width is chosen over the full real or imaginary integer array and is used for all its frequency bands, rather than assigning 16, 32 and 64 bits to different radial intervals.

*Spatial Residual with a Safety Margin and Decoding*

The stored integers define reconstructed Fourier coefficients $\tilde{F}$ and a spatial prediction $p$, where $i$ is the imaginary unit and $F_r^{-1}$ denotes the inverse real FFT:

$$\widetilde{F}(k)=\Delta_F\left[q_R(k)+i\,q_I(k)\right],\ p=f32\left(F_r^{-1}(\widetilde{F})+\mu\right).$$

The encoder computed the integer spatial residual $r(v)$ using a residual tolerance $\varepsilon_r = 0.99\varepsilon$ and quantization interval $2\varepsilon_r = 1.98\varepsilon$. The decoded volume, denoted $y_0$ in the expression below, was formed directly from this residual and the prediction:

$$r(v)=rint\left(\frac{f64(x(v))-f64(p(v))}{1.98\,\varepsilon}\right),$$

$$y_0(v)=f32\left[f64(p(v))+1.98\,\varepsilon\,f64(r(v))\right].$$

In exact arithmetic, nearest-integer quantization at spacing $1.98\varepsilon$ limits the residual error relative to the working input to $0.99\varepsilon$. The remaining $0.01\varepsilon$ provides headroom for working-array and final-output casts; it is not a universal floating-point error bound. Float64 subtraction and residual addition reduce intermediate cancellation and rounding before the required float32 output cast; they do not add information to the input. The final decoded RI is the residual-corrected output $y_0$.

Every reported error was measured against the native reference using full-volume float64 differences. All primary HT settings passed at $\varepsilon = 10^{-4}$. The precision sweep identified two failing settings at a smaller tolerance (Fig. 4a); these are explicitly marked and excluded from bound-compliant performance claims. Supplementary Methods S1 and S3 describe the numerical limitations. Paired measurements of the residual-margin change are described in Supplementary Methods S7.

Each packet for a single field contains the mean, coefficient streams, residual and metadata needed for decoding without the original volume. An additional voxel replacement step (a guard), when used, stores voxel indices and source values to replace decoded values that fail the error check. Complete packet sizes include all headers, compressed arrays, any enclosing container, and replacement data required by a comparator or the joint tensor error limits. Packet framing is documented with the reproduction materials in Supplementary Methods S10.

*Reference Codecs and Parameter Selection*

BitRound–Pcodec and zfp were evaluated on the same float32 working inputs. BitRound used the Numcodecs mantissa-rounding implementation[15,16] followed by Pcodec at level 6. zfp provides a compressed representation of multidimensional floating-point arrays[23,39]; we used zfpy in fixed-accuracy mode with its header enabled and verified the final error against the native reference.

For each comparator, we used the tested setting that produced the smallest complete packet while keeping the RI error within the specified limit at every voxel. We reused these settings and the Fourier coefficient quantization rule in the residual-margin experiments without further tuning. The main figure therefore compares these fixed settings; other settings may yield smaller packets. Supplementary Methods S2 lists the tested parameter values and reports two additional comparisons: one that stores replacement values for voxels exceeding the error limit, and another that compares codecs at similar measured maximum errors. Supplementary Methods S3 reports the precision and coefficient–correction experiments. Supplementary Methods S11 reports packet sizes for FixedScaleOffset–Pcodec at levels 6 and 12; its encoding and decoding times were not measured.

*Metrics, Verification and Figure Preparation*

Compression ratio was $B_{ref} / B_{packet}$, where $B_{ref}$ is the uncompressed reference size and $B_{packet}$ includes all headers and payloads; storage savings relative to a comparator were $1 - B_{proposed} / B_{comparator}$, using the complete packet sizes $B_{proposed}$ and $B_{comparator}$. Reference sizes used 4 bytes per scalar voxel, six unique float32 fields for iDTT and two float32 fields for QAT. Throughput was reported in MiB/s. Independent decoding checks evaluated maximum absolute error, RMSE and the number of bound violations over full volumes. Native and decoded images shared spatial locations and display limits. Complete metric definitions, verification procedures and figure-selection rules are provided in Supplementary Methods S8.

*Timing, Throughput and Memory Measurements*

All codecs were timed on an Apple M4 Max host with 16 logical CPUs and 64 GiB of memory, using one FFT worker per job. Static HT and extension timings covered complete encoding or decoding, including required encoder-side reconstruction and any voxel replacement steps required by the comparator codecs or the tensor error limits, while excluding parameter screening, file I/O and separate quality validation. Time-lapse batch timing included warm local reads and buffered packet writes. Timings used medians from 5 static HT, 9 time-lapse or 21 extension repetitions at each setting. Extension peak resident set size (RSS) was measured separately in fresh processes. Software versions, warmups, timing order and memory protocols are provided in Supplementary Methods S4 and S6.

*Bounded Concurrent Processing of Time-lapse Volumes*

Each time point was encoded separately. Both Fourier and BitRound–Pcodec used a thread pool to process up to $W$ volumes simultaneously, and results were collected in input order. Figure 3 shows how performance changes with $W$ for each codec. The horizontal BitRound–Pcodec line shows its serial performance ($W = 1$). Comparisons between the two codecs at the same $W$, together with timing and memory measurement methods, are reported in Supplementary Methods S4.

*Tensor and Absorption Tomography*

For iDTT, the six unique components of the symmetric relative-permittivity tensor $\epsilon$ were encoded independently, and the full tensor was restored by symmetric duplication. Its isotropic part $m$ and deviatoric part $D$ were defined by $m = \frac{tr(\epsilon)}{3}$, $D = \epsilon - m I$, where tr is the trace and $I$ is the 3 × 3 identity matrix. All codecs satisfied $|\delta m(v)| \leq 10^{-4}$ and $\| \delta D(v) \|_F \leq 3 \times 10^{-5}$ at every voxel, where $\delta$ denotes the compression-induced change and $\| \cdot \|_F$ is the Frobenius norm. After

decoding all components, the same final check was applied to every codec. At voxels that failed either tensor error check, all six values were replaced with their stored source values. For the Fourier method, this joint check followed residual reconstruction without separate replacements of individual component values. Compression ratios used the six unique float32 fields as the uncompressed reference. The $m$/$\boldsymbol{D}$ decomposition defines error metrics here; the packet still stores six unique tensor components, not seven $m$/$\boldsymbol{D}$ fields. In a separate $m$/$\boldsymbol{D}$ representation, trace($\boldsymbol{D}$)=0 leaves only five independent deviatoric components, plus m. Historical representation tests in Supplementary Methods S5 use different error limits applied separately to each tensor component.

For QAT, we encoded the co-registered RI $n$ and extinction coefficient $\kappa$ volumes from the 520 nm melanoma reconstruction. The constant medium RI $n_m$ was subtracted from $n$ before encoding and added back after decoding. The absolute error tolerances were $\varepsilon_n = 10^{-4}$ for $n$ and $\varepsilon_\kappa = 10^{-5}$ for $\kappa$. RI contrast used Fourier coding and spatial residual quantization; $\kappa$ used an FFT-free mean-plus-spatial-residual representation. Both proposed fields used spacing 1.98 times their respective tolerances without additional voxel replacement; the comparator codecs retained a separate replacement step for each field. Data definitions, fixed settings and validation are provided in Supplementary Methods S5. The physical definition is the imaginary part of the complex RI $\tilde{n} = n + i\kappa$, with absorption coefficient $\mu_a = 2k_0\kappa$ and $k_0 = 2\pi/\lambda$. The supplied reconstruction stores this quantity as $-V_{imag}$.

## Results

### *Precision-informed Fourier Quantization with Voxelwise Error Control*

The algorithm compresses a three-dimensional RI volume by quantizing its Fourier coefficients, then corrects local errors using a spatial residual with a 1% safety margin (Fig. 1). Fourier coefficients describe RI variations at different spatial frequencies. Quantization maps nearby coefficient values to shared integer levels and can map weak coefficients to zero, reducing the information needed for lossless coding. The main HT tolerance was $\varepsilon = 10^{-4}$ RI units, selected with reference to our experimental measurements using the HT-X1 system (see Methods); final compliance was checked against the native reconstruction.

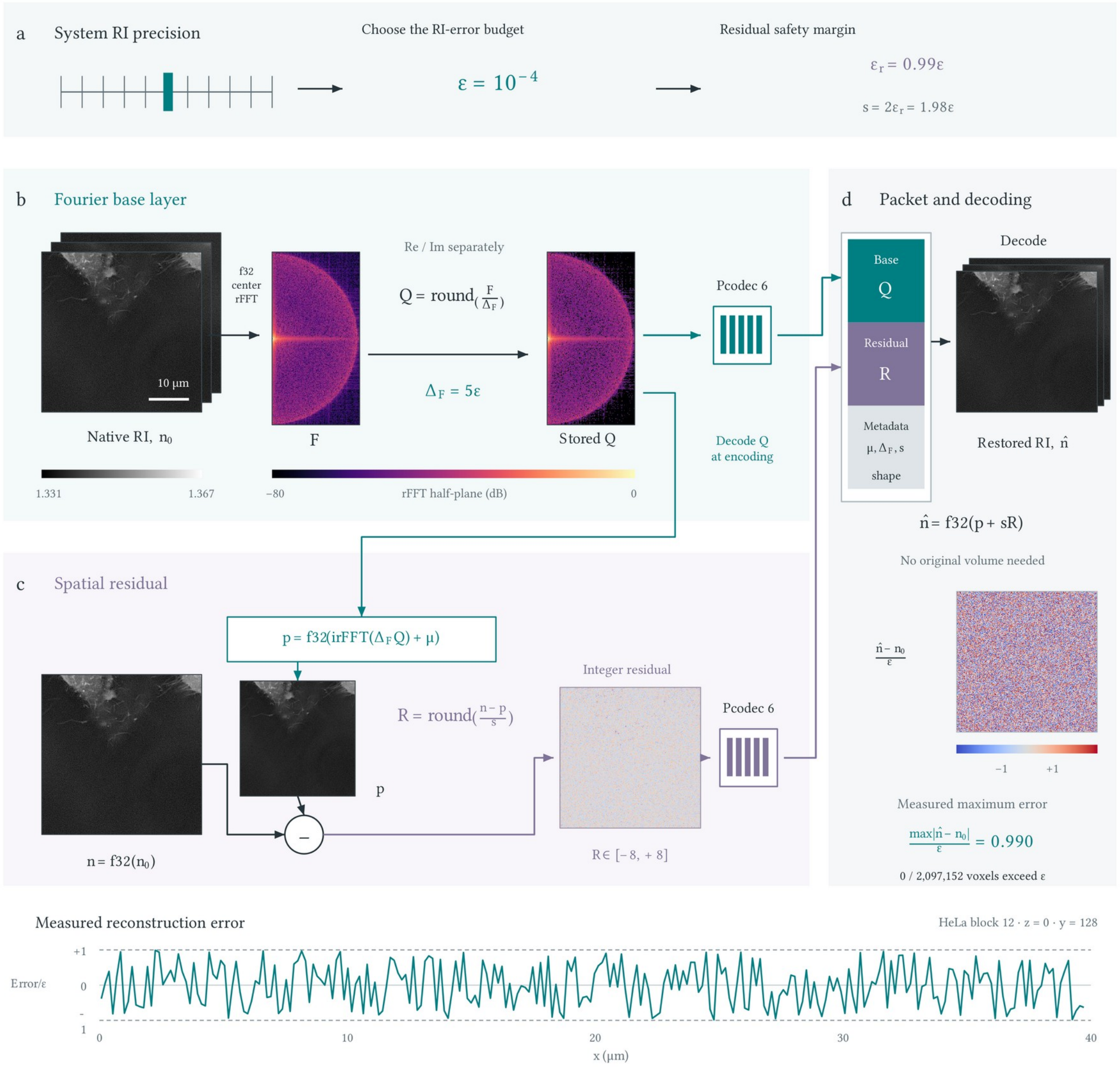

Fig. 1. Precision-informed Fourier compression with a residual safety margin. Refractive index (RI); fast Fourier transform (FFT); $\varepsilon$, assigned absolute RI-error tolerance. (a) The study-selected RI-error tolerance is ε; the residual tolerance is $\varepsilon_r = 0.99\varepsilon$, giving spacing $2\varepsilon_r = 1.98\varepsilon$. (b) A native HeLa subvolume is cast to float32, mean-centered and transformed by a three-dimensional orthonormal real FFT. Real and imaginary coefficients are quantized separately at $\Delta_F = 5\varepsilon$ and losslessly coded. Spectra show the $k_z = 0$, $k_x \geq 0$ half-plane with shifted $k_y$ and a common −80 to 0 dB amplitude reference. (c) Stored coefficients generate the encoder-side prediction; a spatial integer residual is computed from the float32 working input without storing additional voxel replacement values. (d) Coefficients, residual and metadata form a self-contained packet. The restored RI and signed error are shown with an unsmoothed measured error profile below. All 2,097,152 voxels satisfy $\varepsilon = 10^{-4}$ in this example. Paired RI images share display limits. Scale bar, 10 μm. Smaller-tolerance limitations are shown in Fig. 4.

Each complete scalar packet contains the quantized Fourier coefficients, spatial residual and metadata needed to decode the volume without the original tomogram. Real and imaginary

Fourier coefficients of the mean-centered float32 volume were quantized and losslessly encoded with Pcodec. The coefficients defined the Fourier prediction, and an integer residual with spacing 1.98$\varepsilon$ produced the decoded float32 values. No additional voxel indices or replacement values were stored.

Spatial residual quantization reduced the representative HeLa maximum RI error from $1.64\times10^{-3}$ after Fourier quantization to $9.90\times10^{-5}$, or 0.99ε, without additional voxel replacements (Fig. 1). All 2,097,152 voxels satisfied $\varepsilon = 10^{-4}$, and the decoded root-mean-square RI error was $5.72\times10^{-5}$. An independent decoder verified the full three-dimensional subvolume; Supplementary Methods S9 explains how compliance was checked before rounding.

*Smaller Packets for Cellular and Organoid Refractive Index Tomograms*

Fourier coding reduced complete packet sizes in both public RI datasets while satisfying the $10^{-4}$ RI bound (Fig. 2). The comparison used 16 subvolumes per dataset, each containing $32 \times 256 \times 256$ voxels, with compression ratios measured against the same float32 reference size.

Fourier compression achieved aggregate ratios of 7.89 for HeLa and 9.50 for organoid data, compared with 5.55 and 6.80 for BitRound–Pcodec[13,16] and 3.90 and 4.29 for the zfp[23] configurations shown in Fig. 2c. Fourier packets were consequently 29.6% and 28.4% smaller than BitRound–Pcodec packets, and 50.5% and 54.9% smaller than the displayed zfp packets. Native and decoded sections, shown with matched display limits and enlarged regions, retained the visible cellular structures (Fig. 2a,b).

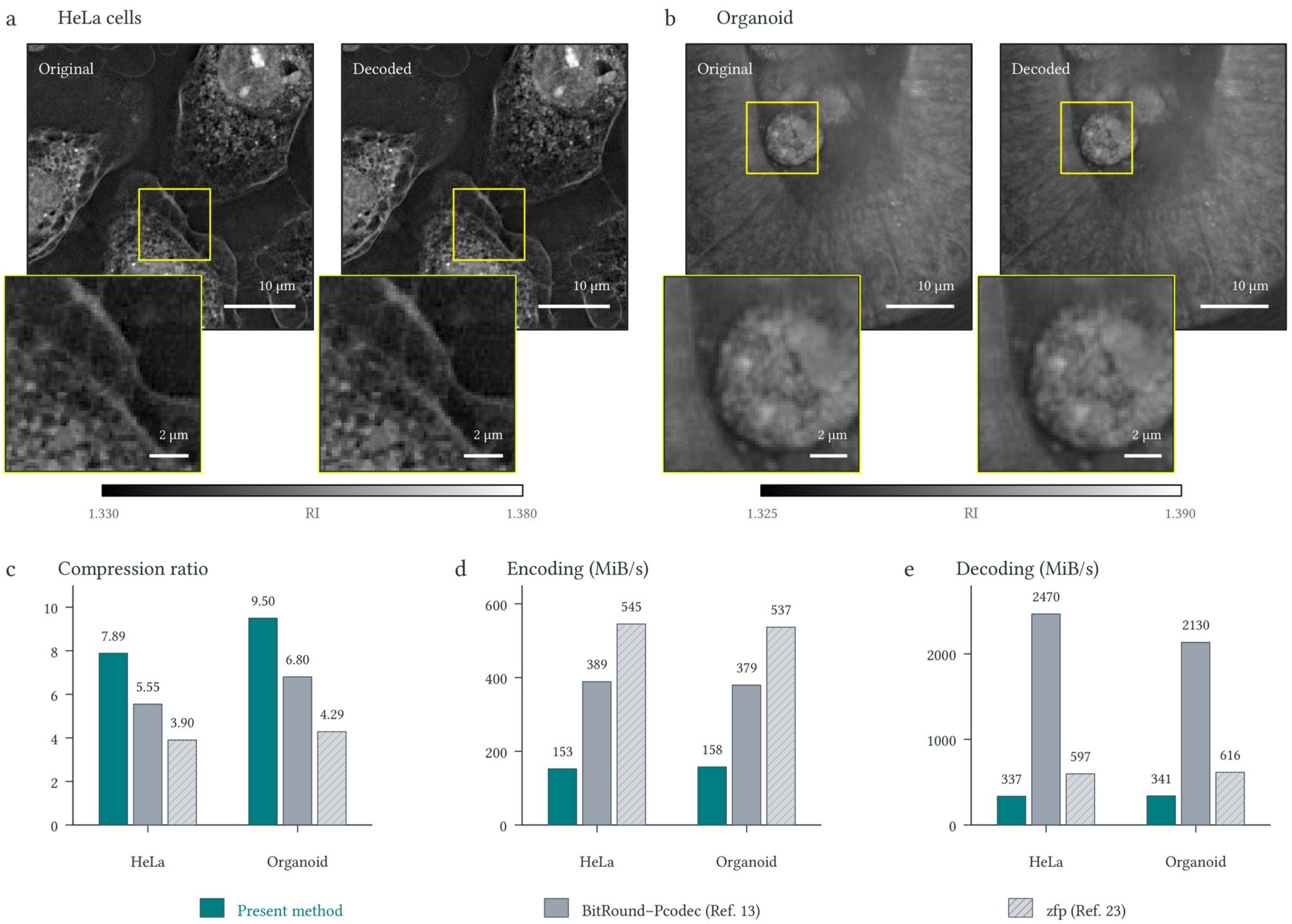

Fig. 2. Storage efficiency and single-volume processing cost with a residual safety margin. (a,b) Native and Fourier-decoded refractive index (RI) sections of HeLa cells and organoids, with matched 64×64-pixel details and identical display limits within each specimen pair. Scale bars, 10 µm in overviews and 2 µm in details. The organoid illustration is an additional cell-rich chunk excluded from the aggregate bars. (c–e) Complete-packet compression ratio, encoding throughput and decoding throughput for 16 benchmark subvolumes per dataset, each containing 32×256×256 voxels. Ratios use a common float32 reference size. Fourier coding uses $\Delta_F = 5\varepsilon$ and residual spacing $1.98\varepsilon$ without additional voxel replacements, with $\varepsilon = 10^{-4}$.

Fourier packets remained smaller than BitRound–Pcodec packets at nearly the same realized maximum RI error. BitRound–Pcodec reached approximately $6.11\times10^{-5}$ RI units, whereas the primary Fourier setting reached approximately $0.99\times10^{-4}$. Recomputing the residual-margin codec with its tolerance set to BitRound's realized maximum error gave ratios of 6.67 and 8.08 for HeLa and organoid data, retaining 16.9% and 15.8% packet-size reductions with no violations. Fourier packets were also smaller in the additional comparator tests with voxel replacement; the settings and verification results are reported in Supplementary Methods S2.

Fourier encoding and decoding were slower than both comparator codecs for single volumes. Encoding throughput was 153–158 MiB/s, compared with 379–389 MiB/s for BitRound–Pcodec and 537–545 MiB/s for the displayed zfp configurations (Fig. 2d). Decoding throughput was 337 –341 MiB/s, compared with $(2.13–2.47)\times10^3$ MiB/s and 597–616 MiB/s, respectively (Fig. 2e). In the same comparison on identical inputs, replacing the earlier $2\varepsilon$ residual followed by voxel replacement with the $1.98\varepsilon$ residual reduced encoding time by 6.29% and 7.42% for HeLa and organoid data.

*Concurrent Processing Reduces Time-lapse Wall Time without Changing Packets*

Concurrent processing shortened the total encoding and decoding times for the 16-volume organoid RI series without changing the packets produced by serial processing (Fig. 3). Consecutive subvolumes from the same spatial location at time indices 75–90 were encoded independently, with the concurrent-volume limit $W$ varied from 1 to 16. Fourier and BitRound–Pcodec retained compression ratios of 8.59 and 6.36, respectively, at every $W$. Packet hashes were unchanged, and all Fourier-decoded voxels remained within the $10^{-4}$ RI-error tolerance.

Increasing W from 1 to 16 reduced total Fourier encoding time from 0.886 to 0.172 s, a 5.14-fold speedup, and decoding time from 0.396 to 0.0823 s, a 4.81-fold speedup. The corresponding amortized encoding times were 55.4 and 10.8 ms per volume, and decoding times were 24.8 and 5.15 ms per volume (Fig. 3b,c). These values are batch wall times divided by 16. The lowest measured medians occurred at $W = 8$ for encoding and $W = 9$ for decoding; $W = 16$ was not uniformly optimal.

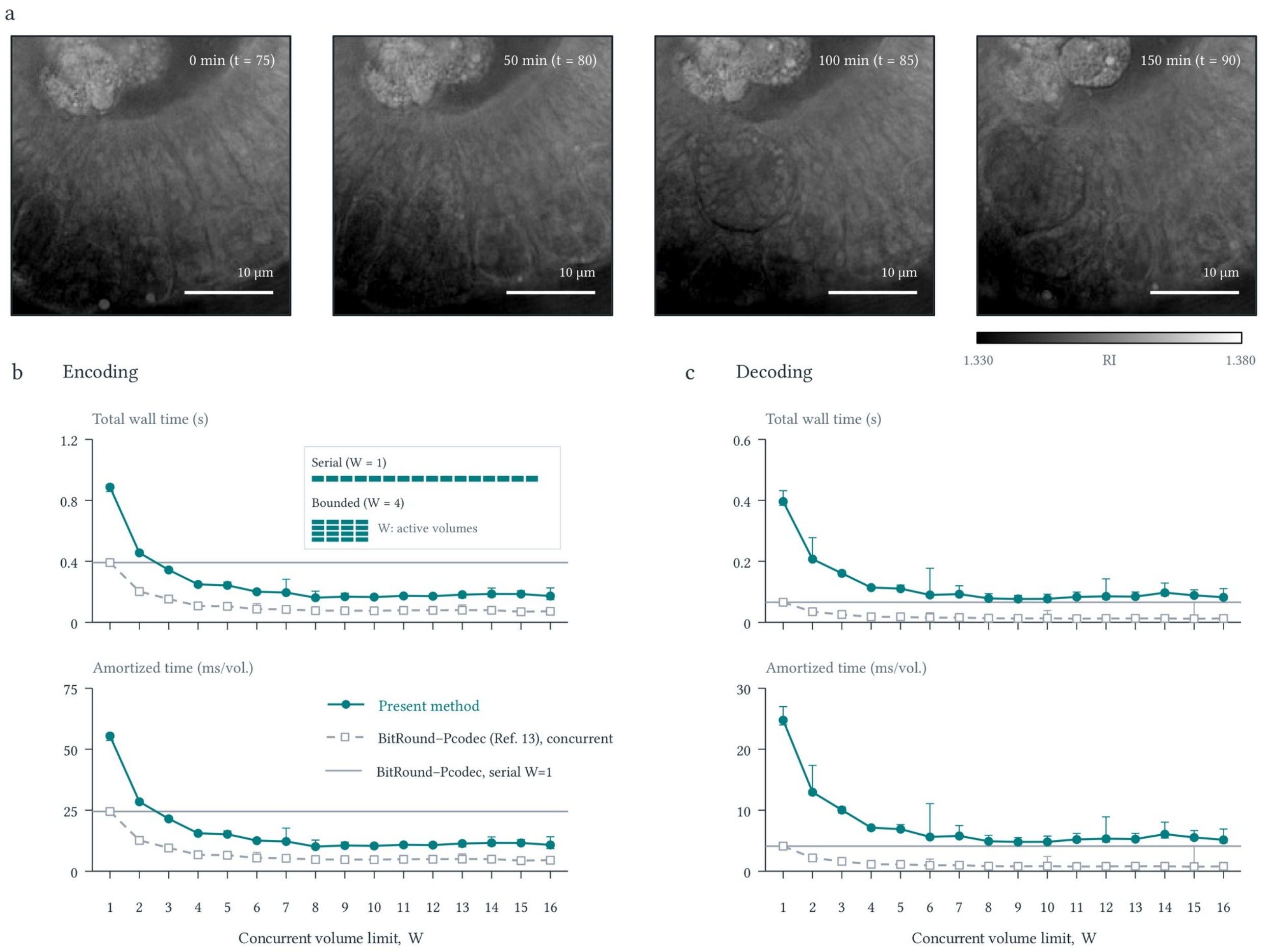

Fig. 3. Bounded concurrent processing of a time-lapse region of interest (ROI) at unchanged compression ratio. Fast Fourier transform (FFT); W, concurrent-volume limit. (a) Native sections of the same organoid ROI at $t$ = 75, 80, 85 and 90, corresponding to 0, 50, 100 and 150 min relative to t75. Images use the same plane, crop and linear grayscale limits. Scale bars, 10 µm. Timing uses the complete 32×256×256 subvolume at each of 16 distinct time points $t$ = 75–90; the schematic illustrates independent-volume scheduling. (b,c) Encoding and decoding wall times at every integer $W$ from 1 to 16, with one FFT worker per Fourier volume. Upper plots show total time for 16 volumes; lower plots show total/16, rather than task latency. Points are medians of nine repeats; error bars span minima to maxima. Fourier uses $\Delta_F = 5\varepsilon$ and residual spacing $1.98\varepsilon$ without additional voxel replacements, with $\varepsilon = 10^{-4}$. Packets are identical across $W$, compression is 8.59× and every decoded voxel stays within the error tolerance relative to its native value. Dashed gray lines show BitRound–Pcodec. Solid gray horizontal lines show the local tolerance-matched BitRound–Pcodec W1 median held fixed across the axis, using the method family of Lee et al.[13].

### *Precision and Coefficient Spacing Determine Different Parts of the Storage Tradeoff*

Reducing the target RI tolerance generally increased packet size in both representative subvolumes (Fig. 4a). Across 14 logarithmically spaced tolerances from $5\times10^{-3}$ to $5\times10^{-6}$, with $\Delta_F = 5\varepsilon$, compression ratios fell from 843 to 3.87 for HeLa and from 847 to 4.22 for organoid data. Of the 28 settings, 26 satisfied their own native-reference tolerance. At $\varepsilon = 8.51\times10^{-6}$, both inputs failed: maximum error was $1.002\varepsilon$, with 607 and 592 violating voxels. Red crosses mark these noncompliant points. Passing again at $5\times10^{-6}$ shows that the behavior cannot be summarized by a single lower tolerance threshold.

Figure 4b compares Fourier prediction alone with two separate residual-quantization settings. Relative to conventional residual quantization at spacing 2ε, the adopted spacing 1.98ε reduced the maximum native-reference error to approximately 0.99003ε in both representative inputs, with zero violations and no additional voxel replacement data.

Coarser Fourier quantization did not continuously reduce total packet size because coefficient savings were offset by a larger spatial residual stream (Fig. 4c,d). At a fixed tolerance of $10^{-4}$, increasing $\Delta_F/\varepsilon$ from 1 to 10 reduced HeLa coefficient storage from 1.02 to 0.282 MiB, while residual storage increased from 0.0329 to 0.708 MiB. The smallest tested complete HeLa packet occurred at $\Delta_F/\varepsilon=5$, giving a compression ratio of 8.26. The organoid minimum occurred at a ratio of 3, giving 9.77; the fixed ratio of 5 gave 9.69. All 16 fixed-tolerance settings passed. Thus, total storage reflected a balance between coefficient and residual bytes, and the best tested spacing differed between inputs.

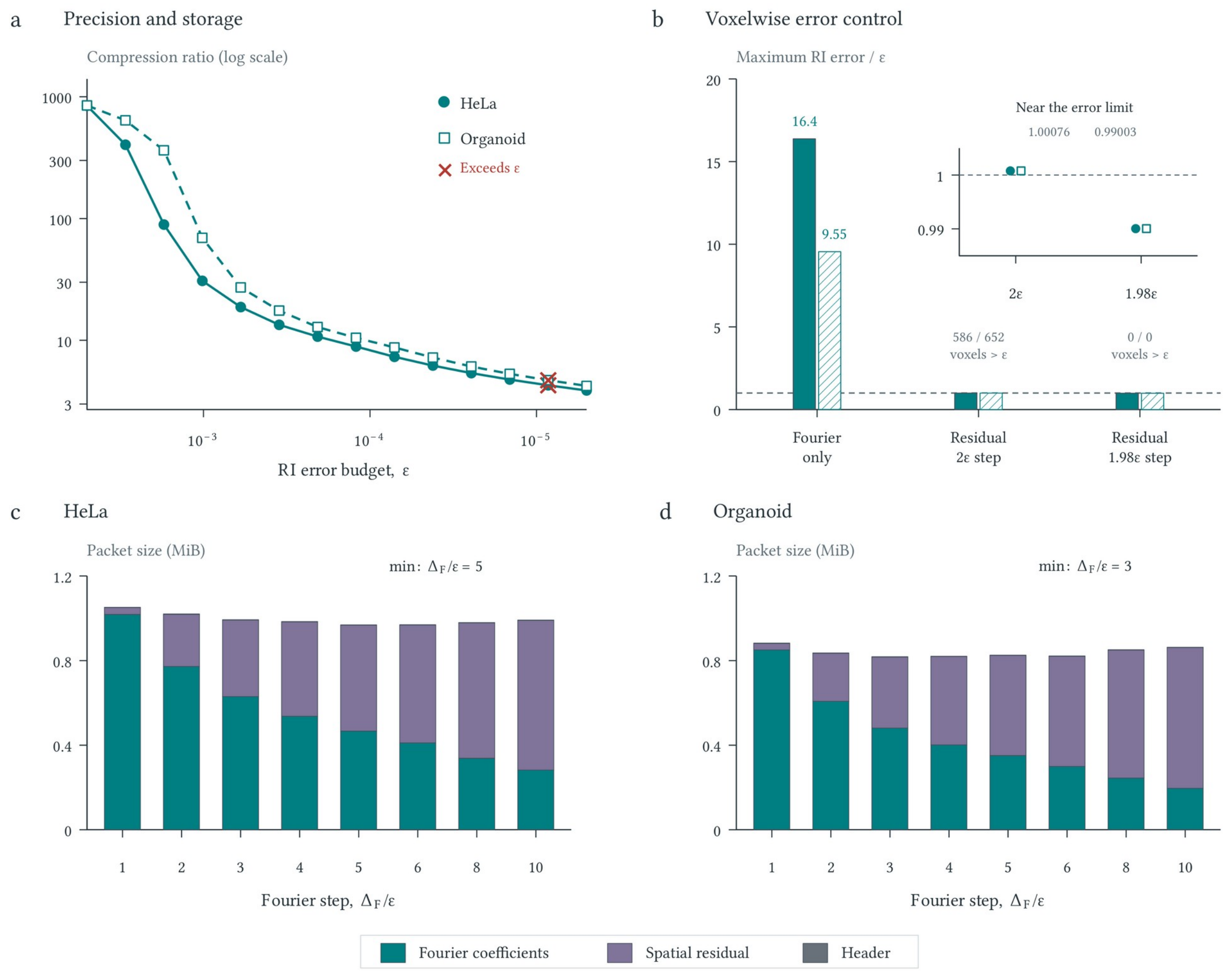

Fig. 4. Precision, residual error control and packet composition. $\varepsilon$, assigned refractive index (RI)-error tolerance; $\Delta_F$, Fourier coefficient interval. (a) Complete-packet compression ratio at 14 logarithmically spaced tolerances per representative subvolume, from $5\times10^{-3}$ to $5\times10^{-6}$, with $\Delta_F = 5\varepsilon$ and residual spacing 1.98ε without voxel replacement. Red crosses identify measured settings exceeding the native-reference bound; they are not bound-compliant performance points. (b) Maximum native-reference RI error divided by $\varepsilon = 10^{-4}$ for Fourier prediction alone, conventional residual quantization at $2\varepsilon$ and the adopted residual quantization at $1.98\varepsilon$. The two residual settings were tested independently. The dashed line marks the limit; the inset reproduces the near-boundary bar heights. Counts give violating HeLa/organoid voxels. (c,d) Coefficient, residual and header bytes for eight $\Delta F/\varepsilon$ ratios at $\varepsilon = 10^{-4}$; all 16 settings pass the fixed tolerance. Compression ratios use the same uncompressed float32 size as the reference.

### *Extension to Tensor and Absorption Tomograms*

The compression scheme also accommodated iDTT tensor components and paired QAT RI/absorption fields under error bounds tailored to their physical quantities[29,30] (Fig. 5). The tests used one symmetric dielectric-tensor ROI and one total variation-regularized melanoma ROI containing paired RI-contrast and extinction-coefficient fields. Each extension enforced error bounds on the physical quantities being stored.

The tensor comparison used six unique components under common isotropic and deviatoric error bounds. These components reconstructed the full symmetric $3 \times 3$ tensor. Letting $m = tr(\epsilon)/3$ and $D = \epsilon - m I$, all codecs were required to satisfy voxelwise $|\delta m| \leq 10^{-4}$ and $\|\delta D\|_F \leq 3 \times 10^{-5}$, with the same final step that replaces all six values at voxels that fail either tensor error check. Compression ratios used the six float32 fields, totaling 48 MiB, so eliminating symmetric duplication did not itself inflate the reported gain.

For the tensor ROI, Fourier compression achieved a ratio of 7.89, compared with 4.17 for BitRound–Pcodec and 4.96 for zfp, reducing packet size by 47.1% and 37.1%, respectively (Fig. 5c). The maximum error in mean relative permittivity and the maximum Frobenius norm of the deviatoric-tensor error were $1.48 \times 10^{-5}$ and $3.00 \times 10^{-5}$. The maximum principal-RI error was $9.66 \times 10^{-6}$, below $10^{-4}$, and the maximum error among pairwise principal-index differences was $1.31 \times 10^{-5}$.

Fourier tensor compression used more processing time and peak memory than BitRound–Pcodec and zfp. Encoding and decoding took 335 and 140 ms, compared with 157/23.3 ms for BitRound–Pcodec and 194/82.4 ms for zfp. Peak process RSS was 298/296 MiB for Fourier encoding/decoding, versus 273/224 MiB and 230/220 MiB for the comparators. All methods used the same final step that replaces all six tensor values at voxels that fail either error check.

QAT compression used separate native-reference error tolerances for the paired RI-contrast $\Delta n$ and extinction-coefficient $\kappa$ fields. We used the total variation-regularized green channel data at 520 nm, with $\Delta n = n - n_{med}$, $n_{med} = 1.336$, and $\kappa = -V_{imag}$. The two co-registered fields were each $32 \times 256 \times 256$ voxels, giving a combined float32 size of 16 MiB. Separate native-reference tolerances of $10^{-4}$ for $\Delta n$ and $10^{-5}$ for $\kappa$ were checked for all three codecs. Proposed fields used $1.98\varepsilon$ residuals without additional voxel replacement; both comparator codecs stored source values to replace decoded voxels that failed the corresponding field’s error check.

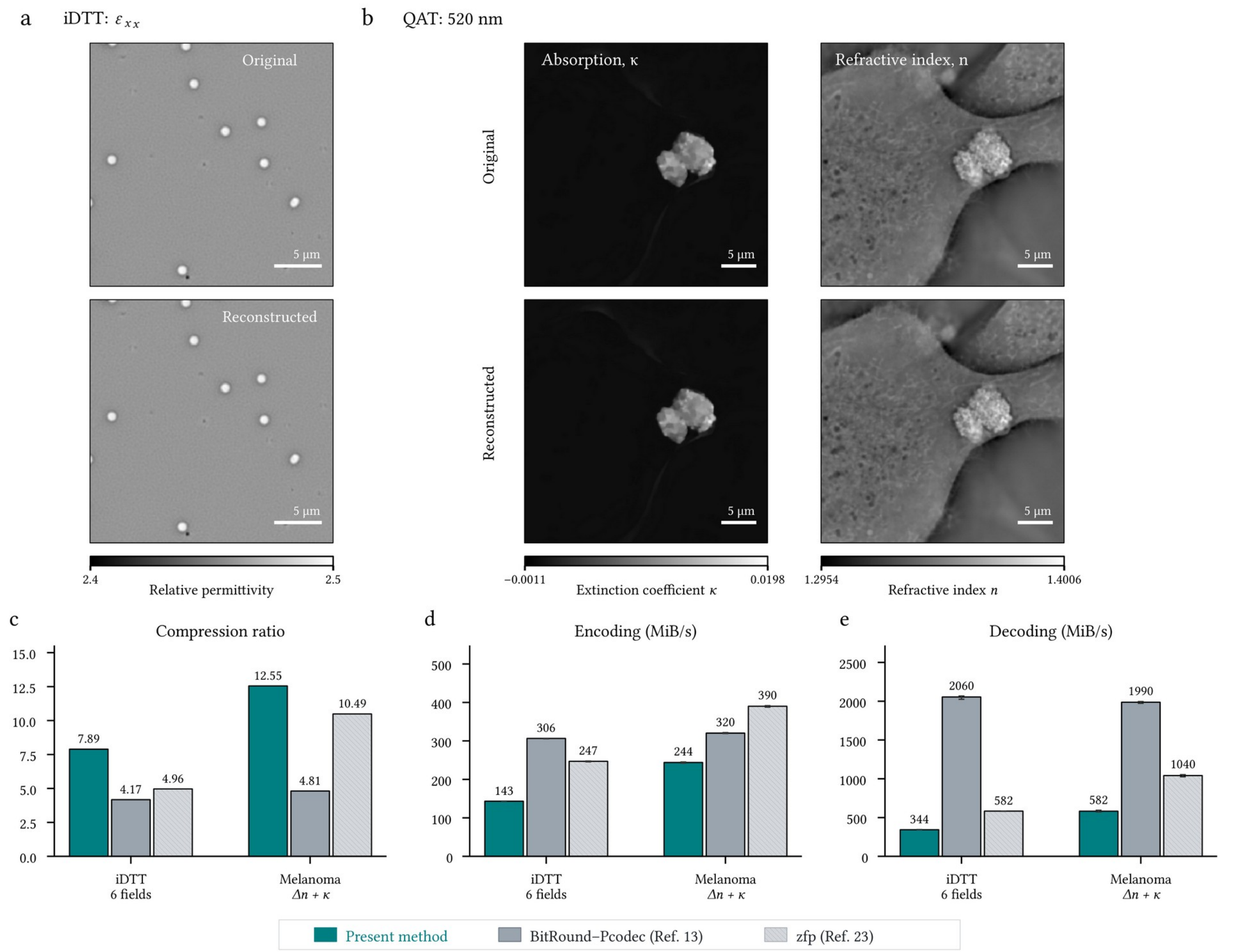

Fig. 5. Extension to tensor permittivity and quantitative absorption tomography (QAT). Incoherent dielectric tensor tomography (iDTT); refractive index (RI). (a) Original and reconstructed relative-permittivity component $\varepsilon_{xx}$ slices from iDTT. The colorbar reports relative permittivity, not RI. (b) Matched original and reconstructed extinction-coefficient κ and RI n slices from the melanoma total variation reconstruction at 520 nm. RI is displayed as the supplied medium index, 1.336, plus $\Delta n$. Scale bars, 5 µm. (c–e) Compression ratio and encoding/decoding throughput for six unique float32 tensor fields and paired $\Delta n/\kappa$ fields. Throughput derives from the median of 21 timed runs after three warmups; error bars span the transformed timing interquartile range. Reference sizes are 48 MiB for iDTT and 16 MiB for QAT. Proposed scalar residuals use spacing 1.98ε without additional replacements of individual component values. The final step that replaces all six values at voxels failing either tensor error check is used for all iDTT methods to satisfy $|\delta m| \leq 10^{-4}$ and $\|\delta \boldsymbol{D}\|_F \leq 3\times10^{-5}$. QAT tolerances are $\varepsilon_n = 10^{-4}$ and $\varepsilon_\kappa = 10^{-5}$; the comparators retain separate voxel replacement for each field. All evaluated volumes satisfy these error limits.

The QAT $\kappa$ path used a constant float32 prediction formed from its float64 mean, followed by spatial residual quantization at $1.98\times10^{-5}$ and Pcodec coding. This avoids an FFT for $\kappa$ while retaining its spatial variation in the residual. This choice followed tests in which all quantized $\kappa$ Fourier coefficients were zero. The selected representation and coefficient spacing were held fixed for the residual-margin evaluation (Supplementary Methods S5).

The combined Fourier and mean-plus-residual QAT representation produced smaller packets than both comparators, with greater runtime and peak memory use in the tested melanoma ROI. The combined compression ratio was 12.6, compared with 4.81 for BitRound–Pcodec and 10.5 for zfp, corresponding to 61.7% and 16.4% smaller packets (Fig. 5c). Maximum errors were $9.90\times10^{-5}$ for $\Delta n$ and $9.90\times10^{-6}$ for $\kappa$, with no violations. Encoding/decoding took 65.5/27.5 ms, compared with 50.0/8.05 ms for BitRound–Pcodec and 41.0/15.4 ms for zfp (Fig. 5d,e). Peak RSS was 176/158 MiB, compared with 169/93.9 MiB and 125/88.6 MiB. The QAT coefficient spacing and the representation used for each field were selected on this same ROI and kept unchanged for the margin experiment. Supplementary Methods S5 reports the separate RI-contrast and extinction-coefficient packet sizes and the common container overhead.

## Discussion

Direct compression of the reconstructed three-dimensional RI volume reduced storage with every tested primary-setting voxel within its RI-error ceiling. The method combines a Fourier representation of the RI volume with spatial residual correction and native-reference verification. At $10^{-4}$ RI units, HeLa and organoid packets were approximately 28–30% smaller than BitRound–Pcodec packets; tightening the Fourier tolerance to BitRound's realized error retained a 15.8–16.9% reduction (Fig. 2). The latter comparison shows that the observed storage gain persists when the methods use nearly the same realized maximum error. This gain comes with additional single-volume computation, making storage capacity and transfer bandwidth important factors in codec selection[13].

Hologram codecs store holograms or complex wavefields for reconstruction, whereas the present method directly stores the reconstructed three-dimensional RI field. These hologram-domain codecs[40] use approaches such as Fourier-coefficient reduction[41], nonuniform companding[42], bandwidth-aware image-plane coding[43], assessment through reconstructed RI volumes[44], and adaptive short-time Fourier quantization[45]. Here, the error criterion applies directly to the decoded RI voxels. Single-cell RI tomograms have also been encoded using three-dimensional Zernike descriptors[46]. Related precedents include three-dimensional Fourier quantization for volume compression[47] and magnetic resonance imaging compression that limits signal-to-noise ratio loss by relating error variance to thermal noise[48]. These precedents motivate direct volumetric coding with an RI-unit fidelity constraint for quantitative HT storage.

A reconstructed RI volume can be described by Fourier coefficients that specify the amplitude and phase of RI variations at different spatial scales. The numerical precision retained in a stored RI volume can be reduced subject to a specified reconstruction-error tolerance. Quantizing the Fourier coefficients reduces this numerical precision by mapping nearby coefficient values to shared integer levels, which can be encoded efficiently. Here, our experimental measurements using the HT-X1 system informed the selected tolerance; the compression experiments assess numerical fidelity relative to the saved reconstruction. Since coefficient errors combine during the inverse transform, spatial residuals use a tighter tolerance and final native-reference checks establish compliance for each evaluated setting. The method thus stores the reconstructed three-dimensional RI field with fewer bits while keeping the compression error within the study-selected tolerance. In HT, accessible spatial bandwidth also determines direction-dependent optical resolution[49]. The present codec retains all half-spectrum positions rather than applying a hard optical-bandwidth mask.

Most of the storage reduction achieved came from quantization at the selected RI-error tolerance, followed by lossless compression. Direct spatial quantization already provided substantial compression under the same error limit. Fourier-domain quantization provided a further reduction in stored size of 9.3% for HeLa and 14.2% for organoid volumes under matched lossless compression settings (7.8% and 12.3% relative to Pcodec level 12; Supplementary Methods S11), supporting its use in the present codec. In these same inputs, 55.7% of HeLa and 70.5% of organoid quantized Fourier scalar coefficients were zero, while 89.2% and 92.7% of residual integers were −1, 0 or +1. Grouping the same coefficient integers by frequency band also reduced their encoded size (Supplementary Methods S12). These measurements support the compact coefficient-plus-residual representation for the tested inputs.

The resulting advantage reflects the combined effects of the data structure, quantization, and encoding strategy, and its magnitude may vary across datasets.

Compressing the reconstructed volume lets us control the error in the RI values used for analysis. The three-dimensional Fourier coefficients describe spatial variation in the quantitative RI field, and a full-volume native-reference check determines whether a selected setting meets the tolerance. Memmolo et al. encoded single-cell RI tomograms using three-dimensional Zernike descriptors[46]. General volume compression has previously used three-dimensional Fourier quantization[47], magnetic resonance imaging compression has linked error variance to thermal noise and image signal-to-noise ratio[48], and SPERR and the FFCz preprint provide transform-correction precedents[26,27]. Our contribution is the integration and evaluation of volumetric Fourier coding with a study-selected RI tolerance and native-reference voxelwise validation in HT, including explicit tests of the margin's numerical limits. This differs from compressing two-dimensional holographic wavefields and assessing their effects after RI reconstruction[44]. A bound on the decoded volume controls the stored quantitative output directly, without requiring hologram decoding and tomographic reconstruction to recover that output.

Compression efficiency depends on how much data are stored in the Fourier coefficients and how much are needed for spatial correction. Coarser quantization reduces the information stored in the Fourier coefficients but leaves larger deviations for the spatial residual to correct. Finer quantization preserves a more accurate Fourier prediction at the cost of additional coefficient storage. The minima observed in Fig. 4c,d identify an intermediate spacing at which these competing contributions yield the smallest complete packet. Their different locations show that the best tested balance depends on the input. Effective compression therefore requires choosing Fourier precision to minimize coefficient and correction storage together; the adopted residual margin controls quantization error and native-reference validation identifies compliant settings.

A residual tolerance of $0.99\varepsilon$ eliminated the extra pass that identified voxels exceeding the single-field error limit and stored their indices and replacement values at the primary HT setting, reducing encoding time in the paired comparison. The 1% margin is an empirical numerical allowance. At $\varepsilon = 8.51\times10^{-6}$, input and output casts together exhausted the margin and produced small violations in both precision-sweep inputs (Fig. 4a). Deployment at another tolerance, numerical library or architecture therefore requires a check of every decoded voxel against the original reconstruction or a separately validated rounding strategy. The final iDTT joint physical correction remains necessary for the joint isotropic and deviatoric tensor error limits.

Bounded concurrent processing shortened the time needed to encode or decode independent volumes without changing the compressed data. Independent chunk processing is consistent with chunked storage workflows; the present timing experiment does not establish end-to-end OME-Zarr integration. Across the 16-volume organoid series, W1→W16 reduced encoding and decoding wall times by factors of 5.14 and 4.81, while packet bytes, the compression ratio of 8.59 and decoded fidelity were unchanged (Fig. 3). Limiting active volumes permits processing without holding the complete acquisition in memory. The lowest medians occurred at W8 and W9 for encoding and decoding, respectively. These results describe the tested host and finite batch; detailed scheduling behavior and matched-concurrency comparisons are given in Supplementary Methods S4.

The coding framework accommodated scalar RI volumes, iDTT and QAT, with error criteria tailored to the physical quantities being stored. For iDTT[29], separate tolerances constrained errors in mean permittivity and in the traceless part of the tensor that describes dielectric anisotropy. These trace- and Frobenius-based criteria are invariant under orthogonal changes of basis, although component-wise coding can yield basis-dependent packet sizes. Using the six independent tensor components as the uncompressed reference excluded gains from symmetric duplication. For QAT[30], the framework accommodated co-registered RI contrast and extinction coefficient $\kappa$ under separate component-specific error tolerances. The RI contrast retained a volumetric Fourier representation, whereas all quantized Fourier coefficients of the extinction coefficient vanished at the selected spacing. The selected mean-plus-residual representation avoids this redundant transform. The spatial residual retained the extinction coefficient's variation around its mean, so a constant prediction was sufficient for this field at the selected tolerance. Under the matched physical error criteria used in each experiment, the proposed method produced smaller complete packets than BitRound–Pcodec and zfp for both extensions, including a 16.4% reduction relative to zfp for the tested QAT ROI, at the cost of greater runtime and peak memory use (Fig. 5). The κ result supports a mean-plus-residual representation at the selected tolerance; it does not demonstrate that Fourier coding benefits the absorption field.

Validation should extend beyond voxelwise reconstruction error to the preservation of quantitative analysis results. The bound limits numerical changes relative to the saved reconstruction and does not establish physical RI accuracy. Accuracy in derived quantities is a separate objective, addressed explicitly in multilevel scientific-data reduction[53]. Dry-mass metrology[54] and quantitative organoid segmentation[55] offer task-level tests of compressed RI data. Broader validation should include calibrated precision measurements and tests of integrated RI, segmentation and birefringence. Principal-axis accuracy also requires separate assessment near tensor eigenvalue degeneracy. Artificial intelligence-based segmentation, cell-type classification and virtual staining may tolerate compression errors larger than the $10^{-4}$ RI-unit tolerance used here. Characterizing the precision required by each task could enable more efficient compression under validated task-specific fidelity criteria. Such validation should encompass the complete analysis pipeline, because dry-mass precision depends on optical-path measurements, background estimation and cell segmentation[56].

Downstream validation should compare RI-derived lipid concentration, dry mass and threshold-based lipid-droplet segmentation between native and compressed volumes[57]. Lymphocyte-subtype classification provides a further endpoint based on morphological and biochemical features extracted from RI tomograms[58]. For virtual staining, validation should assess both inferred compartment masks and the dry-mass estimates obtained by applying those masks to quantitative phase data[59].

Further work could test discrete cosine and wavelet transforms, assess the contribution of the Fourier transform and four-band partition, and explore frequency-dependent error allocation using calibrated noise estimates and voxelwise error validation. Future work may also test whether the observed compression gains translate to large-scale imaging workflows that include storage, transfer and analysis. Runtime and peak RSS were measured on one host, and whether decoding on different processor architectures meets the same error limits remains untested. The HTPack implementation stores a checksum of the decoded array and rejects a nonmatching reconstruction. A checksum mismatch identifies a difference from the certified decoded array; it does not by itself establish that the RI-error tolerance has been exceeded. Longer streaming benchmarks should include reads from uncached storage, data transfer, checks that decoding on different platforms meets the error limits and concurrency selection based on available CPU and memory resources. Integration with chunked bioimaging formats would allow evaluation of the complete storage-to-analysis workflow[12].

## Conclusion

Precision-informed three-dimensional Fourier coding reduced complete-packet storage for the tested RI tomograms under a verified $10^{-4}$ voxelwise error bound. Packets were 28–30% smaller than BitRound–Pcodec packets, with a 15.8–16.9% reduction retained after matching its realized maximum error. The fixed margin failed at one smaller tolerance in both precision-sweep inputs, so its validity requires numerical checks. Additional computation accompanied the primary-setting gain; concurrent processing improved batch throughput while preserving identical packets. Initial iDTT and QAT experiments extended the framework to component-specific error criteria.

## Disclosures

Y.K.P. has financial interests in Tomocube, a company that commercializes HT instruments. Juyeon Park receives financial support from Tomocube.

## Code and Data Availability

The public HeLa and organoid RI datasets used for the static and time-lapse HT comparisons are available from the BioImage Archive under accession S-BIAD1739 (https://doi.org/10.6019/S-BIAD1739). The archive members, subvolume selection rules and ROIs used in this study are specified in the Methods and Supplementary Methods. The iDTT and QAT reconstruction data used for the extension experiments will be deposited in the BioImage Archive; accession numbers will be added when available. The research implementation and HTPack command-line interface will be released on GitHub.

**Acknowledgments**

This work was supported by the National Research Foundation of Korea (NRF), funded by the Ministry of Science and ICT (MSIT), Republic of Korea (Grant Nos. RS-2024-00442348 and RS-2022-NR068141); the Korea Institute for Advancement of Technology (KIAT) through the International Cooperative R&D Program (Grant No. P0028463); the Bio & Medical Technology Development Program of the NRF, funded by MSIT (Grant No. RS-2024-00440577); and the Commercialization Promotion Agency for R&D Outcomes (COMPA), funded by MSIT (Grant No. 2710110765). YongKeun Park also acknowledges support from the Merck fellowship.
We thank Hervé Hugonnet for his insightful comments and constructive feedback. OpenAI Codex was used to assist with language editing and manuscript formatting.

**Author Contributions**

YongKeun Park: conceptualization, methodology, supervision, funding acquisition, and writing—original draft. Dohyeon Lee: data curation, software, formal analysis, validation, visualization, and writing—review and editing. Juheon Lee: methodology, resources, validation, and writing—review and editing. Sehyeon Lee: resources, validation, and writing—review and editing. Juyeon Park: project administration, supervision, and writing—review and editing.

## Author Biographies

Dohyeon Lee conducts research in biomedical optics at KAIST. His interests include holotomographic data processing, scientific data compression, and computational methods for managing large three-dimensional quantitative imaging datasets.

Juheon Lee conducts research in biomedical optics at KAIST. His research focuses on dielectric tensor tomography, polarization-resolved imaging, and quantitative three-dimensional characterization of optical anisotropy.

Sehyeon Lee conducts research in biomedical optics at KAIST. His interests include computational microscopy, quantitative absorption tomography, and reconstruction of three-dimensional optical properties from intensity measurements.

Juyeon Park conducts research in the Biomedical Optics Laboratory at KAIST. Her research focuses on developing label-free three-dimensional imaging and deep-learning methods for quantitative analysis of biological tissues and organoids. In particular, she investigates virtual staining of holotomographic images, with the goal of understanding cellular organization and dynamics without fluorescent labeling.

YongKeun Park is an endowed chair professor in the Department of Physics at KAIST and director of the KAIST Virtual 3D Biology Center. His research spans biomedical optics, wave physics, and holotomography, with a particular focus on developing label-free 3D imaging and quantitative analysis technologies for biological cells, tissues, and advanced 3D culture models. He is a fellow of Optica and SPIE.

# Supplementary Information

## Precision-informed Fourier compression of refractive index tomograms with voxelwise error control

**Dohyeon Lee[a,b], Juheon Lee[a,b], Sehyeon Lee[a,b], Juyeon Park[a,b], YongKeun Park[a,b,c*]**

[a]Department of Physics, Korea Advanced Institute of Science and Technology (KAIST), Daejeon 34141, Republic of Korea

[b]KAIST Institute for Imaging, KAIST, Daejeon 34141, Republic of Korea

[c]Tomocube Inc., Daejeon 34109, Republic of Korea

*Correspondence: YongKeun Park, yk.park@kaist.ac.kr

Reference numbers correspond to the bibliography in the main article.

### S1 Data Selection and Error Reference

We used full-resolution HeLa and organoid refractive index (RI) arrays. Each dataset contributed 16 randomly selected subvolumes of $32 \times 256 \times 256$ voxels; incomplete boundary chunks were excluded. The input lists and source hashes identify the exact data used.

The saved HeLa and organoid float64 arrays were the error references, and their float32 casts were the working inputs. Residual spacing $1.98\,\varepsilon$ limits exact-arithmetic quantization error to $0.99\,\varepsilon$. Input and output rounding also contribute to the final error, so we checked every decoded voxel against the native reference. Proposed single-field packets stored no extra voxel replacement values. The float32 working arrays and the 4-byte-per-voxel compression-ratio denominator are distinct from this native reference. We selected the primary holotomography (HT) tolerance of $\varepsilon = 10^{-4}$ RI units with reference to our experimental measurements using an HT-X1 system (Tomocube Inc., Republic of Korea). The tests reported here establish compression fidelity relative to the saved reconstructions; they do not establish the absolute accuracy or measurement precision of those reconstructions.

### S2 Comparator Settings and Additional Comparisons

For BitRound, we tested 11-18 retained mantissa bits. For zfp, we tested tolerances from $2 \times 10^{-5}$ to $1.6 \times 10^{-3}$. We selected the smallest total packet size for which every voxel met the $10^{-4}$ RI-error limit. The selected settings were 13 bits for BitRound and tolerance $4 \times 10^{-4}$ for zfp; neither needed voxel replacement. These settings and the Fourier rule $\Delta_F = 5\,\varepsilon$ were then kept fixed.

We also tested coarser BitRound and zfp settings with voxel replacement. Their packet sizes included both voxel indices and source values. At 12 retained bits, BitRound ratios were 4.12 and 4.56 for HeLa and organoid. At zfp tolerance $1.6 \times 10^{-3}$, ratios were 4.77 and 5.34. All met the native-reference error limit, but their packets were larger than the selected BitRound packets.

To compare methods at nearly the same measured maximum error, we set the Fourier tolerance to the selected BitRound maximum for each dataset, approximately $6.11 \times 10^{-5}$. Fourier ratios were 6.67 and 8.08, giving 16.9% and 15.8% smaller packets for HeLa and organoid. All 32 packets passed independent decoding. Parameter searches were excluded from timing; exact grids and selected values are retained in the benchmark records.

**S3 Precision and Fourier-spacing Experiments**

Figure 4 used one fixed HeLa and one fixed organoid subvolume, selected by their native RI interquartile range. We tested 14 logarithmically spaced error tolerances from $5 \times 10^{-3}$ to $5 \times 10^{-6}$ with $\Delta_F = 5\varepsilon$. At approximately $\varepsilon = 8.51 \times 10^{-6}$, both inputs failed: maximum errors were $1.002\varepsilon$, with 607 and 592 violating voxels. The other 26 settings passed, including the smaller endpoint. Thus the tests do not define a single lower tolerance threshold below which all inputs fail. At the failing HeLa voxel, input casting, residual quantization and final output casting contributed approximately $-5.95\times10^{-8}$, $-8.412\times10^{-6}$ and $-5.14\times10^{-8}$ RI units, respectively. Their sum exceeded the assigned tolerance because the combined cast errors exhausted the $8.51\times10^{-8}$ safety margin. Alignment of the quantization grid with float32 values explains why a smaller tested tolerance could pass.

At $\varepsilon = 10^{-4}$, we tested coefficient-spacing ratios $\Delta_F / \varepsilon$ of 1, 2, 3, 4, 5, 6, 8 and 10. All 16 packets passed. Figure 4c,d separates coefficient, residual and header storage; the minima are the smallest packets among these tested settings. Figure 4b compares the Fourier prediction with two separately tested residual spacings, $2\varepsilon$ and $1.98\varepsilon$.

The fixed rule $\Delta_F = 5\varepsilon$ was chosen before the benchmark. All half-spectrum positions were retained. The four radial bands used the same coefficient spacing and integer width; separating bands changed only how values were grouped for lossless compression. Coefficients were stored in flattened C-order within each band.

**S4 Time-lapse Processing**

Figure 3 used 16 consecutive organoid subvolumes from the same spatial region at archive time indices 75-90. Each volume was encoded independently. The scheduler kept at most $W$ jobs submitted but not yet consumed, collected results in input order and admitted the next job after consuming a result. The full series did not need to be held in memory.

Each codec, stage and concurrency setting was measured nine times after one warmup. Codec and concurrency order were reversed on alternating repetitions. Encoding included local input reads, float32 casting, compression and buffered packet writes. Decoding included packet reads and reconstruction. Downloading, decoded-array file writes and separate quality checks were excluded. Scheduler overhead was included.

Figure 3 shows batch wall time and that time divided by 16. The latter is amortized time per volume, not individual job latency. The fixed BitRound line is the local $W = 1$ median. BitRound remained faster at matched $W$; at $W = 16$, its amortized encoding and decoding times were 4.52 and 0.796 ms per volume. Packet hashes and decoded arrays were unchanged across concurrency settings.

**S5 Tensor and Absorption Data**

*Dielectric-tensor Compression*

The incoherent dielectric tensor tomography (iDTT) input was a real, symmetric relative-permittivity tensor[29]. We encoded six unique components, each containing $32 \times 256 \times 256$ float32 voxels, and restored the remaining components by symmetric duplication. Sampling was 0.0994 micrometers laterally and 0.4 micrometers axially. Compression ratios used the six unique fields, totaling 48 MiB. The input was the e array in 2_DT.mat from the 20260123/ps1um acquisition, with source-array crop [9:41, 222:478, 222:478, :, :]. Figure 5a displays the relative-permittivity component, whereas Figure 5b displays RI and extinction coefficient.

For tensor $\epsilon$, we defined the isotropic part $m$ and deviatoric part $D$ as

$$m=\frac{tr(\epsilon)}{3}, D=\epsilon - m I .$$

All codecs had to satisfy $|\delta m(v)| \le 10^{-4}$ and $\| \delta D(v) \|_F \le 3 \times 10^{-5}$ at every voxel. With decoded-minus-native component errors $e_{ab}$, we evaluated

$$\| \delta D \|_F^2 = \sum_{a \in \{x,y,z\}} (e_{aa} - \delta m)^2 + 2(e_{xy}^2 + e_{xz}^2 + e_{yz}^2).$$

After decoding, all codecs used the same joint check: if either tensor error reached or exceeded its limit within a $10^{-12}$ margin, the packet supplied all six source values at that voxel. This affected 791 voxels for the proposed method. Fourier components used the $1.98\varepsilon$ residual without separate component-wise replacement.

Diagonal and off-diagonal settings were selected before adopting the residual margin and then kept fixed. Selected Fourier scalar tolerances were $1.5 \times 10^{-5}$ and $10^{-5}$; BitRound retained 16 and 4 bits; zfp tolerances were $4.8 \times 10^{-4}$ and $1.2 \times 10^{-4}$. After decoding, we checked symmetry, positive eigenvalues, principal indices and pairwise index differences. These validation calculations were excluded from codec timing.

*Earlier Tensor-representation Comparison*

An earlier test used the previous codec with voxel replacement and a componentwise error limit of $10^{-4}$. On the same region of interest (ROI), isotropic/deviatoric coding produced 12.3% smaller packets than six-component coding, but encoding and decoding were 2.4% and 4.8% slower and peak resident set size (RSS) increased by approximately 48 MiB. These results used different error criteria from Figure 5.

*Refractive Index and Extinction-coefficient Compression*

The quantitative absorption tomography (QAT) input was the co-registered melanoma RI and extinction-coefficient reconstruction at 520 nm, with total variation regularization and no aberration correction[30]. The two fields each contained $32 \times 256 \times 256$ float32 voxels, totaling 16 MiB, with isotropic sampling of 0.135 micrometers. The supplied medium RI was $n_{med} = 1.336$.

We subtracted it before encoding RI and added it back after decoding; the final RI error was checked against the native RI.

The selected error tolerances were $\varepsilon_n = 10^{-4}$ and $\varepsilon_\kappa = 10^{-5}$. BitRound retained 6 and 3 bits for RI contrast and extinction coefficient; zfp tolerances were $1.6 \times 10^{-3}$ and $1.6 \times 10^{-4}$. Both comparators used voxel replacement where needed. Fourier settings were selected on the same ROI before adopting the residual margin and were not searched again.

The selected RI-contrast coefficient spacing was approximately $4.22 \times 10^{-4}$. For $\kappa$, all quantized Fourier coefficients were zero. We therefore omitted its fast Fourier transform (FFT) and used the native mean, calculated in float64 and cast to float32, as a constant prediction. Pcodec compressed the residual quantized at $1.98 \times 10^{-5}$, without extra voxel replacement.

The proposed RI-contrast packet occupied 0.862 MiB and the extinction-coefficient packet 0.412 MiB. Each includes its own field header and compressed streams; the common container added 661 bytes. The complete packet was 1.27 MiB, giving the combined ratio of 12.6 relative to the 16 MiB input. For BitRound–Pcodec, the RI-contrast and extinction-coefficient packets occupied 2.12 and 1.21 MiB; the corresponding zfp values were 1.2 and 0.322 MiB. The combined proposed result therefore includes Fourier coding of RI contrast and FFT-free mean-plus-residual coding of the extinction coefficient. For this ROI, the proposed extinction-coefficient packet was larger than the zfp packet; the smaller combined packet resulted from the reduction in RI-contrast storage.

The supplied reconstruction defines $\kappa = -V_{imag}$, the negative imaginary reconstruction field. Figure 5b uses the same native and decoded arrays as the performance measurements, at the fixed plane $z = 10$. Each pair shares linear grayscale limits. Negative reconstructed $\kappa$ values were retained and are not interpreted as negative physical absorption. Exact source fields, crops and settings are recorded with the input manifests.

## S6 Timing and Memory Measurements

Experiments ran on an Apple M4 Max with 16 logical CPUs and 64 GiB of memory. Each Fourier job used one FFT worker. Tensor and QAT measurements used one outer job and one thread for OpenMP and BLAS. Software versions are recorded with the benchmarks.

Static HT measurements used one warmup and five encode/decode repetitions per input and setting. Tensor and QAT measurements used three warmups and 21 repetitions. Timings included complete codec operations: prediction, residual coding, headers and any required comparator or joint tensor replacement. Inputs were already in memory; file I/O, parameter searches and separate quality checks were excluded. Time-lapse timing included the local reads and packet handling described in S4.

Tensor and QAT peak RSS was the maximum of three fresh-process measurements, including inputs, outputs and runtime memory. Time-lapse RSS was sampled in the benchmark process. These two memory protocols were not treated as equivalent.

## S7 Effect of the Residual Safety Margin

We compared the previous residual spacing $2\varepsilon$ followed by voxel replacement with the adopted spacing $1.98\varepsilon$ without replacement. Both used identical inputs, Fourier coefficients and comparator settings.

At $\varepsilon = 10^{-4}$, the new margin reduced encoding time by 6.29% for HeLa and 7.42% for organoid, with zero violations. Packet size increased by 0.124% and decreased by 0.430%, respectively, because the new spacing changed residual compressibility. Simply removing replacement while retaining spacing $2\varepsilon$ left 586 HeLa violations, with maximum error $1.001 \times 10^{-4}$.

## S8 Metrics and Independent Verification

Compression ratio was $B_{raw} / B_{packet}$, and savings relative to a comparator were $1 - B_{proposed} / B_{comparator}$. Packet sizes included all headers, payloads and required replacement data.

Reference sizes used 4 bytes per scalar voxel, six fields for iDTT and two for QAT. Throughput was uncompressed bytes divided by elapsed time, reported in MiB/s, where $1\,MiB = 2^{20}$ bytes.

Independent parsers and decoders checked complete packet lengths, decoded arrays and every voxel against the native reference. We measured maximum absolute error, root-mean-square error, signed mean error and violation counts. The two precision-sweep failures were independently reproduced. Verification used a separately constructed radial-band map but shared FFT and compression libraries with the encoder.

Native and decoded images used the same locations and display limits. Full, unclipped volumes were used for numerical checks. The extra organoid illustration in Figure 2 was not part of its 16-volume benchmark. Data images were not sharpened, denoised or generatively modified. Timing repetitions measured computational variability, not biological replication.

**S9 Numerical Reporting and Error-bound Decisions**

Results are generally reported to three significant figures. Four are used when needed to show a small error-limit violation. Compliance was decided from unrounded decoded differences; a value that rounds to the limit can still lie below it. For example, the iDTT deviatoric maximum in Figure 5 rounds to $3.00 \times 10^{-5}$ but was below that limit. Savings and speedups were also calculated before rounding. Counts, array dimensions and identifiers remain exact; full-precision results and settings are retained in the benchmark records.

**S10 Reproduction Materials and Software Checks**

The reproduction materials contain exact input lists and hashes, fixed codec settings, software versions, benchmark scripts and independent decoders. The HT configuration uses float32 working arrays, native float64 references, an orthonormal real FFT, four radial bands, $\Delta_F = 5\,\varepsilon$, Pcodec level 6 and residual spacing $1.98\,\varepsilon$. Packet-format records specify the header and payload lengths used for complete-size accounting.

The corrected float-RI path in HTPack v0.6.0b1 reproduced the compressed coefficient/residual payloads and decoded arrays for all 32 benchmark inputs, with zero native-reference violations. Its smaller headers changed packet size slightly. Independent decoding confirmed the results. The encoder rejects the two smaller-tolerance failures in Figure 4. These checks do not cover the separate 1,024-chunk cohort used in the earlier coauthor review. The distributed HTPack encoder includes additional certification and checksum operations; its timings are not interchangeable with the research-code timings defined in S6.

**S11 FixedScaleOffset-Pcodec Storage Control**

We tested FixedScaleOffset-Pcodec on the same 32 inputs as Figure 2, using grid spacings $10^{-4}$ and $1.98 \times 10^{-4}$ and Pcodec levels 6 and 12. Packets included filter parameters, shape and framing. All 128 packets met the native-reference $10^{-4}$ error limit without voxel replacement. Runtime and memory were not measured.

At the coarser spacing, compression ratios at levels 6/12 were 7.15/7.27 for HeLa and 8.15/8.33 for organoid. At level 12, Fourier packets were 7.81% and 12.3% smaller, respectively. Maximum errors were approximately $9.91 \times 10^{-5}$. The finer grid gave approximately half the assigned error limit; it was not a comparison at matched measured maximum error.

**S12 Coefficient and Residual Distributions**

We analyzed the same 16 HeLa and 16 organoid inputs used in Figure 2 at the primary tolerance, with the same Fourier coefficient spacing, residual spacing and Pcodec level 6. The complete Fourier packets matched the original packets by SHA-256. Zero-valued quantized real and imaginary Fourier scalars accounted for 55.7% of HeLa and 70.5% of organoid coefficients. Residual integers of −1, 0 or +1 accounted for 89.2% and 92.7%, respectively. These fractions describe the tested representation; zero-valued coefficients were not identified as measurement noise.

Across the 16 inputs in each cohort, Fourier coefficient payloads totaled 8.47 MB for HeLa and 6.17 MB for organoids, while residual payloads totaled 8.52 and 7.93 MB (MB = $10^6$ bytes). Including headers, complete Fourier packets totaled 17.0 and 14.1 MB, compared with 18.8 and 16.5 MB for direct spatial quantization at the same residual grid spacing and Pcodec level. The net storage reductions were 9.28% and 14.2%, after paying for the Fourier coefficients and additional metadata.

To examine grouping, we encoded exactly the same coefficient integers as two real/imaginary streams in FFT order instead of eight band/component streams. Coefficient payloads increased to 9.69 and 6.92 MB. Holding the residual and existing header fixed, this payload substitution reduced the net savings to 2.76% and 9.66%. This comparison changes grouping and ordering together; it is not a universal causal attribution or a measurement of a fully implemented two-stream packet format. The measured benefit depends on the inputs, quantization and entropy coder.